\documentclass[12pt]{article}
\input epsf
\def\be{\begin{equation}}
\def\bea{\begin{eqnarray}}
\def\ee{\end{equation}}
\def\eea{\end{eqnarray}}

\begin{document}
\begin{titlepage}

\begin{flushright}
OHSTPY-HEP-T-00-009\\
hep-th/0005282
\end{flushright}
\vfil

\begin{center}
{\Large {\bf Mesonic Spectrum of Two Dimensional Supersymmetric Theories.}}\\
\ \newline
\ \newline
{\bf Oleg Lunin, Stephen Pinsky}\\
\ \newline
\it{Department of Physics \\ Ohio State University, Columbus, Ohio 43210}
\end{center}

\begin{abstract}
\noindent
We consider a bound state problem for a family of supersymmetric gauge
theories with fundamental matter. These theories can be obtained by a
dimensional reduction of supersymmetric QCD from three dimensions to $1+1$ and
subsequent truncation of some of the fields. We find that the models without
adjoint scalar converge to well-defined continuum limits and calculate the
resulting spectra of these theories. We also find the critical value of
coupling at which an additional massless state is observed. By contrast,
the models containing adjoint scalars, seem to have a continuous mass spectrum
in the limit of infinite volume.
\end{abstract}


\end{titlepage}

\section{Introduction.}

One of the most important problems in Quantum Field Theory is the study
of the bound state spectra of non-abelian gauge theories. There are
several approaches to this problem. For QCD like theories lattice gauge
theory is probably the most popular approach since the approximation does
not break the most important symmetry, gauge symmetry. Similarly for
supersymmetric theories, supersymmetric DLCQ (SDLCQ)is probably the most
powerful approach since the approximation does not break the most
important symmetry, supersymmetry. In this paper will consider
supersymmetric theories and follow this latter approach.

Long ago 't Hooft \cite{thooft} showed that two dimensional models can be
a powerful laboratory for studying the bound state problem. These models
remain popular to this day since they are easy to solve and share many
properties with their four dimensional cousins, most notably stable bound
states. Supersymmetric two dimensional models are particularly attractive
since they are also super-renormalizable. Given that the dynamics of
gauge field is responsible for strong interaction and the formation
of bound states, it comes as no surprise  that a great deal of effort has gone
into investing bound states of pure glue in supersymmetric models
\cite{sakai,kutasov,lpreview}. While such theories capture the essential
properties of the mass spectrum and some of them are relevant for the
string theory \cite{tay98}, the wavefunctions are quite different from
the ones for mesons and baryons. Extensive study of the meson spectrum of
non--supersymmetric theories has been done (see \cite{bpp98} for a
review), but the problem has not been  addressed in a context of
supersymmetric models. In this paper we will introduce seven new two
dimensional supersymmetric models that have not been previously studied
that are particularly useful for studying mesons within a two dimensional
supersymmetric laboratory. Throughout this paper we use a word ``meson'' to
indicate the group structure of the state. Namely we define a meson as a bound
state whose wavefunction can be written as a linear combination of parton
chains, each chain starts and ending with a creation operator in fundamental
representation. In supersymmetric theories the
states, defined this way, can have either bosonic or fermionic statistics.

To simplify the calculation we will consider only the large $N$ limit
\cite{thooft}, which has proven to be  a powerful approximation for bound
state calculations. While baryons can be constructed in this limit
\cite{WittBar}, they have an infinite number of  partons and thus
practical calculations for such states are complicated.  In this paper we
concentrate our attention on the mesonic spectrum. Note that throughout
this paper we completely ignore the zero mode problem \cite{zm5}, however
it is clear that considerable progress on this issue could be made
following our earlier work on the zero modes of the two dimensional
supersymmetric model with only adjoint fields \cite{alptzm}.

The paper has the following organization. In section \ref{SectDefin} we
consider  the three dimensional SQCD and dimensionally reduce it to
$1+1$. We perform the light-cone quantization of the resulting theory by
applying canonical commutation relations at fixed $x^+$ and choosing the
light-cone gauge  ($A^+=0$) for the vector field. After solving the
constraint equations we end up with a model containing $4$ dynamical
fields. We construct the supercharge for the dimensionally reduced theory
and observe that it can also be used to define models with less
supersymmetry.  In particular, in section \ref{SecStat} we study the
mesonic spectrum of  systems without dynamical quarks. We find that one
of these systems (we call  it $A\lambda$) has many light states in SDLCQ,
which probably give rise to a  continuous spectrum in the continuum
limit. The other system, containing only dynamical gluinos, seems not to
have a well-defined bound state problem: all masses are pushed to
infinity in the continuum limit.  In section \ref{SecNoGauge} we study
the systems which do not  include the adjoint scalar. They all share the
same properties: a well-defined continuum spectrum and the existing of a
critical value of the  coupling constant at which the lowest mass bound
state becomes massless. Finally, in section \ref{SecGauge} we study the
remaining theories which include the adjoint  boson and at least one of
the fundamental fields. We find that all this models have a continuous
spectrum, which seems to be a general property of two dimensional
supersymmetric systems with adjoint scalars \cite{alp2}.

\section{Supersymmetric Systems with fundamental matter.}
\label{SectDefin}
We consider the family of supersymmetric models in two dimensions which can
be obtained as the result of dimensional reduction of SQCD$_{2+1}$ and
possible truncation of some fields in the resulting two dimensional theory.
Our starting point is the three dimensional action:
\bea\label{action}
S&=&\int d^3x\mbox{tr}\left(-\frac{1}{4}F_{\mu\nu}F^{\mu\nu}+
\frac{i}{2}{\bar\Lambda}
\Gamma^\mu D_\mu \Lambda +D_\mu \xi^\dagger D^\mu \xi+
i{\bar\Psi} D_\mu\Gamma^\mu\Psi-\right.\nonumber\\
&-&\left.g'\left[{\bar\Psi}\Lambda\xi+
\xi^\dagger{\bar\Lambda}\Psi\right]\right).
\eea
This action describes the system of a gauge field $A_\mu$ and its superpartner
$\Lambda$, both taking values in the adjoint representation of $SU(N)$, and
two complex fields, a scalar $\xi$ and a Dirac fermion $\Psi$, transforming
according to the fundamental representation of the same group. Thus in matrix
notation the covariant derivatives are given by:
\be
D_\mu\Lambda=\partial_\mu\Lambda+ig[A_\mu,\Lambda],\quad
D_\mu\xi=\partial_\mu\xi+ig'A_\mu\xi,\quad
D_\mu\Psi=\partial_\mu\Psi+ig'A_\mu\Psi.
\ee
The action (\ref{action}) is invariant under supersymmetry transformations
which are
parameterized by a two--component Majorana fermion $\varepsilon$:
\bea
\delta A_\mu=\frac{i}{2}{\bar\varepsilon}\Gamma_\mu\Lambda,\qquad
\delta\Lambda=\frac{1}{4}F_{\mu\nu}\Gamma^{\mu\nu}\varepsilon,\nonumber\\
\delta\xi=\frac{i}{2}{\bar\varepsilon}\Psi,\qquad
\delta\Psi=-\frac{1}{2}\Gamma^\mu\varepsilon D_\mu\xi.
\eea
We introduced the commutator of the Dirac matrices:
$$\Gamma^{\mu\nu}=\frac{1}{2}\left[\Gamma^{\mu}\Gamma^{\nu}\right].$$
Using standard techniques one can evaluate the Noether current corresponding
to these transformations:
\bea\label{sucurrent}
{\bar\varepsilon}q^\mu&=&\frac{i}{4}{\bar\varepsilon}\Gamma^{\alpha\beta}
\Gamma^\mu\mbox{tr}\left(\Lambda F_{\alpha\beta}\right)+
\frac{i}{2}D^\mu\xi^\dagger\
{\bar\varepsilon}\Psi+\frac{i}{2}\xi^\dagger{\bar\varepsilon}\Gamma^{\mu\nu}
D_\nu\Psi\nonumber\\
&-&\frac{i}{2}{\bar\Psi}\varepsilon D^\mu\xi+\frac{i}{2}D_\nu
{\bar\Psi}\Gamma^{\mu\nu}\varepsilon\xi.
\eea
We will consider the reduction of this system to two dimensions, which means
that the field configurations are assumed to be independent of the space--like
dimension $x^2$. In the resulting two dimensional system we will implement
light--cone quantization, which means that initial conditions as well as
canonical commutation relations will be imposed on a light--like surface
$x^+=const$. In particular we construct the supercharge by integrating the
current (\ref{sucurrent}) over the light--like surface:
\bea\label{sucharge}
{\bar\varepsilon}Q&=&\int dx^-dx^2\left(
\frac{i}{4}{\bar\varepsilon}\Gamma^{\alpha\beta}
\Gamma^+\mbox{tr}\left(\Lambda F_{\alpha\beta}\right)+
\frac{i}{2}D_-\xi^\dagger\
{\bar\varepsilon}\Psi+\frac{i}{2}\xi^\dagger{\bar\varepsilon}\Gamma^{+\nu}
D_\nu\Psi\right.\nonumber\\
&-&\left.\frac{i}{2}{\bar\Psi}\varepsilon D^+\xi+\frac{i}{2}D_\nu
{\bar\Psi}\Gamma^{+\nu}\varepsilon\xi\right).
\eea
Since all fields are assumed to be independent of $x^2$, the integration over
this coordinate gives just a constant factor, which we absorb by a field
redefinition.

If we consider a specific representation for the Dirac matrices in three
dimensions:
\be
\Gamma^0=\sigma_2,\qquad \Gamma^1=i\sigma_1,\qquad \Gamma^2=i\sigma_3,
\ee
then the Majorana fermion $\Lambda$ can be chosen to be real, it is also
convenient to write the fermions in the component form:
\be
\Lambda=\left(\lambda,{\tilde\lambda}\right)^T,\qquad
\Psi=\left(\psi,{\tilde\psi}\right)^T,\qquad
Q=\left(Q^+,Q^-\right)^T
\ee
In terms of this decomposition the superalgebra has explicit $(1,1)$ form:
\be\label{sualg}
\{Q^+,Q^+\}=2\sqrt{2}P^+,\qquad \{Q^-,Q^-\}=2\sqrt{2}P^-,\qquad
\{Q^+,Q^-\}=0.
\ee
The traditional way of solving the bound state problem \cite{bpp98} is based on
a simultaneous diagonalization of the momentum $P^+$ and the 
Hamiltonian $P^-$, but as
one can see from the structure of (\ref{sualg}), the same problem can be
solved by diagonalizing $P^+$ and $Q^-$ instead \cite{sakai}.

In order to solve the bound state problem we impose the light cone gauge
($A^+=0$), then the supercharges are given by:
\bea\label{Qplus}
Q^+&=&2\int dx^-\left(\lambda\partial_-A^2+
\frac{i}{2}\partial_-\xi^\dagger\psi-\frac{i}{2}\psi^\dagger\partial_-\xi-
\frac{i}{2}\xi^\dagger\partial_-\psi+\frac{i}{2}\partial_-\psi^\dagger\xi
\right)\\
Q^-&=&-2\int dx^-\left(-\lambda\partial_-A^-+
i\xi^\dagger D_2\psi-iD_2\psi^\dagger\xi+\frac{i}{\sqrt{2}}
\partial_-({\tilde\psi}^\dagger\xi-\xi^\dagger{\tilde\psi})\right).\nonumber\\
\eea
Note that apart from  a total derivative these expressions involve only
left--moving components of fermions ($\lambda$ and $\psi$). In fact in the
light--cone formulation only these components are dynamical. To see this we
consider the equations of
motion that follow from the action (\ref{action}), in the light cone gauge.
Three of them serve as constraints rather than as dynamical statements:
\be\label{constraint}
\partial^2_-A^-=gJ,\quad J=i[A^2,\partial_-A^2]+
\frac{1}{\sqrt{2}}\{\lambda,\lambda\}-ih\partial_-\xi\xi^\dagger+
ih\xi\partial_-\xi^\dagger+\sqrt{2}h\psi\psi^\dagger,
\ee
\bea
&&\partial_-{\tilde\lambda}=-\frac{ig}{\sqrt{2}}
\left([A^2,\lambda]+ih\xi\psi^\dagger-ih\psi\xi^\dagger\right),\\
&&\partial_-{\tilde\psi}=-\frac{ig'}{\sqrt{2}}A^2\psi+
\frac{g}{\sqrt{2}}\lambda\xi.
\eea
We introduced the relative coupling for the fundamental matter: $h=g'/g$.
Apart from the zero mode problem \cite{zm5}, one can invert the first
constraint to write the auxiliary field $A^-$ in terms of physical degrees of
freedom. Substituting the result into the expression for the supercharge and
omitting the boundary term, we get:
\be\label{Qminus}
Q^-=-2\int dx^-\left(gJ\frac{1}{\partial_-}\lambda+
g'\xi^\dagger A^2\psi+g'\psi^\dagger A^2\xi\right).
\ee
This supercharge gives rise to a whole family of supersymmetric theories.
Namely one can see that the expression (\ref{Qminus}) is meaningful even if
we exclude some of the fields from the theory. As soon as we have at least
one fermion and at least one field in the adjoint representation,
(\ref{Qminus}) defines an interacting theory with supersymmetry. Some of these
theories were studied before (namely the pure adjoint systems with
\cite{sakai} or without \cite{kutasov} bosons), but many of them are new. In
this paper we will study the mesonic spectrum of all these models, their field
content is summarized in the table \ref{ContentTable}.
\begin{table}
\begin{center}
\caption{Interacting supersymmetric models with fundamental matter.\newline
The numbers refer to the section in the paper where appropriate model is
studied, and symbol --- appears when the models which do not exist.}
\label{ContentTable}
\  \newline
\begin{tabular}{|c|c|c|c|c|}
\hline
            &no fundamentals&    $\psi$&    $\xi$&    $\psi\xi$\\
\hline
$\lambda$&\ref{SecStat}&\ref{SecNoGauge}&\ref{SecNoGauge}&\ref{SecNoGauge}\\
\hline
$A^2$&   ---&  ---&  ---&  \ref{SecGauge}\\
\hline
$\lambda A^2$&\ref{SecStat}&\ref{SecGauge}&\ref{SecGauge}&\ref{SecGauge}\\
\hline
\end{tabular}
\end{center}
\end{table}

In order to solve the bound state problem we apply the methods of
Supersymmetric DLCQ. Namely we compactify the two dimensional theory on a
light--like circle ($-L<x^-<L$), and impose periodic boundary conditions on
all physical fields. This leads to the following mode expansions:
\bea
A^2_{ij}(0,x^-)&=&\frac{1}{\sqrt{4\pi}}\sum_{k=1}^{\infty}\frac{1}{\sqrt{k}}
\left(a_{ij}(k)e^{-ik\pi x^-/L}+a^\dagger_{ji}(k)e^{ik\pi x^-/L}\right),\\
\label{expandLambda}
\lambda_{ij}(0,x^-)&=&\frac{1}{2^{\frac{1}{4}}\sqrt{2L}}\sum_{k=1}^{\infty}
\left(b_{ij}(k)e^{-ik\pi x^-/L}+b^\dagger_{ji}(k)e^{ik\pi x^-/L}\right),\\
\xi_i(0,x^-)&=&\frac{1}{\sqrt{4\pi}}\sum_{k=1}^{\infty}\frac{1}{\sqrt{k}}
\left(c_i(k)e^{-ik\pi x^-/L}+{\tilde c}^\dagger_{i}(k)e^{ik\pi x^-/L}\right),\\
\label{expandPsi}
\psi_{i}(0,x^-)&=&\frac{1}{2^{\frac{1}{4}}\sqrt{2L}}\sum_{k=1}^{\infty}
\left(d_{i}(k)e^{-ik\pi x^-/L}+{\tilde d}^\dagger_{i}(k)e^{ik\pi x^-/L}\right).
\eea
We drop the zero modes of the fields. Including them
could lead to new and interesting effects (see \cite{alptzm}, for example), but
this is beyond the scope of this work.
In the light--cone formalism one treats $x^+$ as a time direction, thus
the commutation relations between fields and their momenta are imposed on
the surface $x^+=0$.
For the system under consideration this means:
\bea\label{CanComRelField}
\left[A_{ij}^2(0,x^-),\partial_-A_{kl}^2(0,y^-)\right]&=&
i\left(\delta_{il}\delta_{kj}-\frac{1}{N}\delta_{ij}\delta_{kl}\right)
\delta(x^--y^-),\\
\left\{\lambda_{ij}(0,x^-),\lambda_{kl}(0,y^-)\right\}&=&\sqrt{2}
\left(\delta_{il}\delta_{kj}-\frac{1}{N}\delta_{ij}\delta_{kl}\right)
\delta(x^--y^-),\\
\left[\xi_i(0,x^-),\partial_-\xi_j(0,y^-)\right]&=&
i\delta_{ij}\delta(x^--y^-),\\
\left\{\psi_{i}(0,x^-),\psi_{j}(0,y^-)\right\}&=&\sqrt{2}
\delta_{ij}\delta(x^--y^-).
\eea
These relations can be rewritten in terms of creation--annihilation operators:
\be
\left[a_{ij},a^\dagger_{kl}\right]=
\left(\delta_{il}\delta_{kj}-\frac{1}{N}\delta_{ij}\delta_{kl}\right),\quad
\left\{b_{ij},b^\dagger_{kl}\right\}=
\left(\delta_{il}\delta_{kj}-\frac{1}{N}\delta_{ij}\delta_{kl}\right),
\ee
\be
\left[c_{i},c^\dagger_{j}\right]=\delta_{ij},\quad
\left[{\tilde c}_{i},{\tilde c}^\dagger_{j}\right]=\delta_{ij},\quad
\left\{d_{i},{d}^\dagger_{j}\right\}=\delta_{ij} \quad
\left\{{\tilde d}_{i},{\tilde d}^\dagger_{j}\right\}=\delta_{ij}.
\ee
In this paper we will discuss numerical results obtained
  in the large $N$ limit, i.e. we neglect $1/N$ terms in the above expressions.
Although $1/N$ corrections may lead to interesting effects \cite{anp98}, they
are beyond the scope of this work. In the presence of fundamental matter,
however, one can
also define different limits by making the number of flavors comparable with
number of colors \cite{Ven76}. While this is an interesting direction for
a future exploration, in the current paper we concentrate on models with one
flavor and an infinite number of colors.

\section{Mesons involving static (s)quarks.}
\label{SecStat}
We begin our consideration with the simplest models involving only 
adjoint matter.
While the glueball spectrum of such models was studied before
\cite{kutasov,sakai,lpreview}, it might also be interesting to look at the
meson--like states. In the string interpretation of these theories
\cite{dak93} such states would correspond to open strings with freely moving
endpoints. In the QCD language the model corresponds to a system of
interacting gluons and gluinos which is bounded by nondynamical (s)quark and
anti-(s)quark. In the large $N$ limit we will have to consider only a single
(s)quark---anti-(s)quark pair, thus the Fock space is constructed from the
states of the following type:
\be\label{state}
{\bar f}_{i_1} a^\dagger_{i_1i_2}(k_1)\dots
b^\dagger_{i_ni_{n+1}}(k_n)\dots f_{i_p}|0\rangle.
\ee
Here $|0\rangle$ is a vacuum defined by annihilation operators $a_{ij}$ and
$b_{ij}$ and ${\bar f}_i$ and $f_{i_p}$ are sets of c--numbers.

The supercharges for such a system can be constructed by eliminating
fundamental matter from the expressions (\ref{Qplus}) and (\ref{Qminus}):
\bea\label{statquark}
Q^+&=&2\int dx^- \lambda\partial_-A^2,\nonumber\\
Q^-&=&-2g\int dx^-\left(i[A^2,\partial_-A^2]+
\frac{1}{\sqrt{2}}\{\lambda,\lambda\}\right)\frac{1}{\partial_-}\lambda.
\eea
The supercharge $Q^-$ has an interesting property which can be seen from its
mode expansion \cite{alp98a}. Acting on states (\ref{state}), this operator
changes the numbers of bosons $a^\dagger$ by an even number ($0$ or $\pm 2$),
thus one can perform the diagonalization of $P^-$ on
two separate spaces: those containing either even or odd number of bosons.
The second supercharge $Q^+$ makes the situation even more interesting:
it maps
one of these spaces into another and thus leads to the same massive spectra
in both sectors (see \cite{alp98a}). We have studied the mesonic mass
spectrum of $\left(Q^-\right)^2$ at low values of the harmonic resolution $k$
and found no massless states, thus the spectra in two different sectors
are completely identical. Note that this property is not
satisfied for the ``glueball spectrum'': there we found a lot of exact
massless states \cite{alp98a,alp2}.

\begin{figure}
\begin{tabular}{cc}
\epsfxsize=2.5in \epsffile{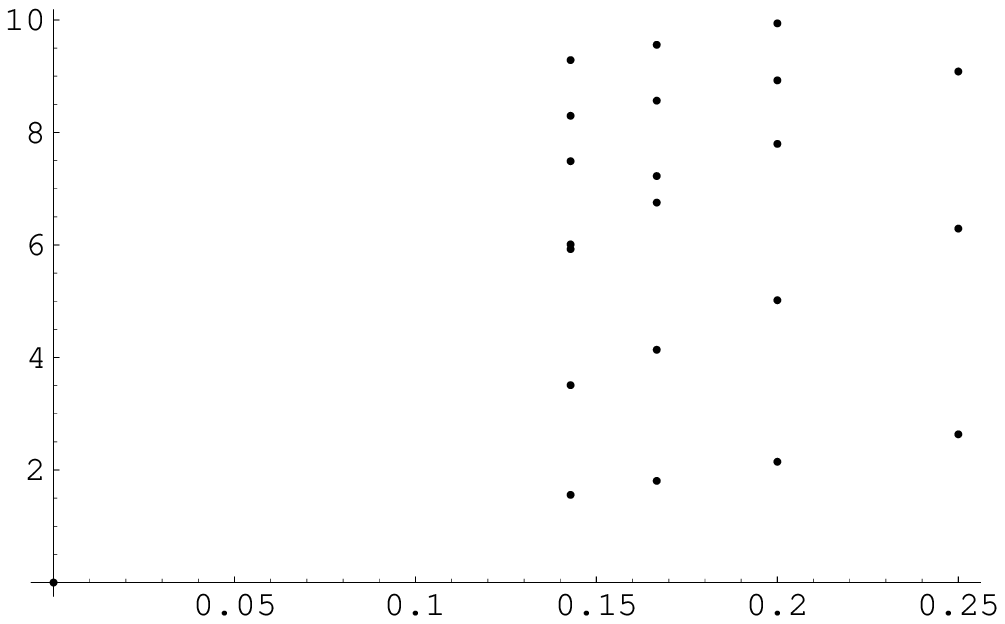}&
\epsfxsize=2.5in \epsffile{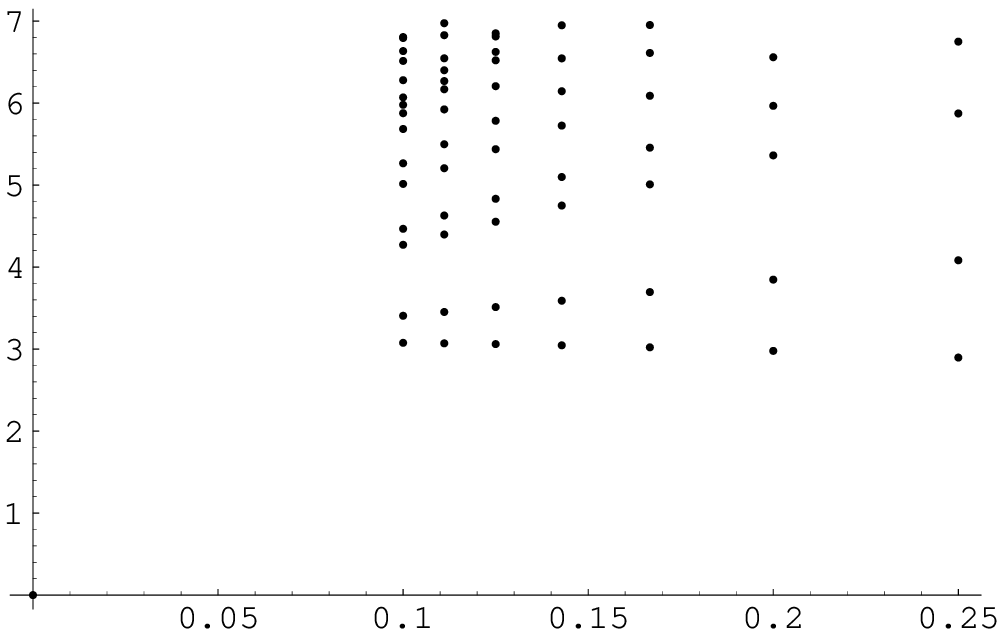}\\
(a)&(b)
\end{tabular}
\caption{(a) Mesonic mass spectrum of $A\lambda$ model in units of
$g_{YM}^2 N /\pi$ as a function of $1/k$.
(b) Eigenvalues of Hamiltonian $P^-=M^2/k$ in units of $g_{YM}^2 N / \pi$
for the mesonic sector of $\lambda$ model. \label{StatFerMeson}}
\end{figure}

In figure \ref{StatFerMeson}a we present the results of the numerical
diagonalization of the Hamiltonian. Note that we considered only one of two
equivalent sectors. Although we have not seen any massless states at any
finite value of resolution (we considered $K=4\dots 7$), it appears that at
least the two lowest states converge to $M=0$ in the continuum limit.
This conclusion is supported by the quadratic fit to the data. The third
lowest state is also well--defined and an extrapolation of its mass gives a
value of $1.83$ for the continuum limit. To reveal the structure of higher
states we need some additional information about their wavefunctions, this will
lead to a clear distinction between nearly degenerate states.

While this direction is definitely worth pursuing, we already can formulate
some interesting properties of the mesonic mass spectrum. Unlike the glueball
case, there are no massless mesons at any finite value of resolution. In the
continuum limit, however, the massless mesons appear, but the number of such
massless states is still an open question. We also see an example of a meson
state converging to a finite mass, and the data presented in Figure
\ref{StatFerMeson}a suggests that there are many such states in the continuum
limit. We should mention that the convergence properties of this model are
very good, which seems to be a general properties of supersymmetric DLCQ as
opposed to the traditional DLCQ approach.
%
%

As we mentioned in the previous section, the supercharge $Q^-$ is
well--defined
as soon as we have at least one fermion. This suggests an interesting
truncation of the supersymmetric adjoint model \cite{kutasov}: one can
eliminate the bosonic field $A^2$ from the theory. This reduces the number of
supersymmetries to $(1,0)$ and the remaining supercharge is given by:
\be
Q^-=-\frac{g}{\sqrt{2}}\int dx^-\left\{\lambda,\lambda\right\}
\frac{1}{\partial_-}\lambda.
\ee
While the ``closed string'' sector of this theory has well--defined continuum
bound states \cite{kutasov,bdk93}, all mesonic masses appear to be
pushed to infinity in the continuum limit. In fact if one looks at the
spectrum of DLCQ Hamiltonian $P^-$, rather than the mass operator $M^2=2KP^-$,
a finite limit can be found. The result, presented in figure
\ref{StatFerMeson}b, has a peculiar property: an extrapolation of the four
lowest eigenvalues gives $3.08$, $3.01$, $3.09$ and $3.06$ accordingly. Thus
it seems that the spectrum of the Hamiltonian has some kind of a threshold.
This
interesting observation, however, does not undermine the fact that this model
does not have any sensible continuum limit.

\section{Models without adjoint scalar.}
\label{SecNoGauge}
As we already saw in the previous section, the simplest supersymmetric model
can be constructed by truncating all fields in the supercharge (\ref{Qminus}),
  except for gluino $\lambda$. In this section we add the dynamics of
fundamental matter to the model. There are three different ways of doing
this, and they give rise to the systems which we call $\lambda\Psi$, 
$\lambda\xi$
and $\lambda\Psi\xi$. We will show that all three systems have a well--defined
mass spectrum and their bound states exhibit similar behavior under a
variation of the coupling constant $h$.

\begin{figure}
\begin{tabular}{cc}
\epsfxsize=2.5in \epsffile{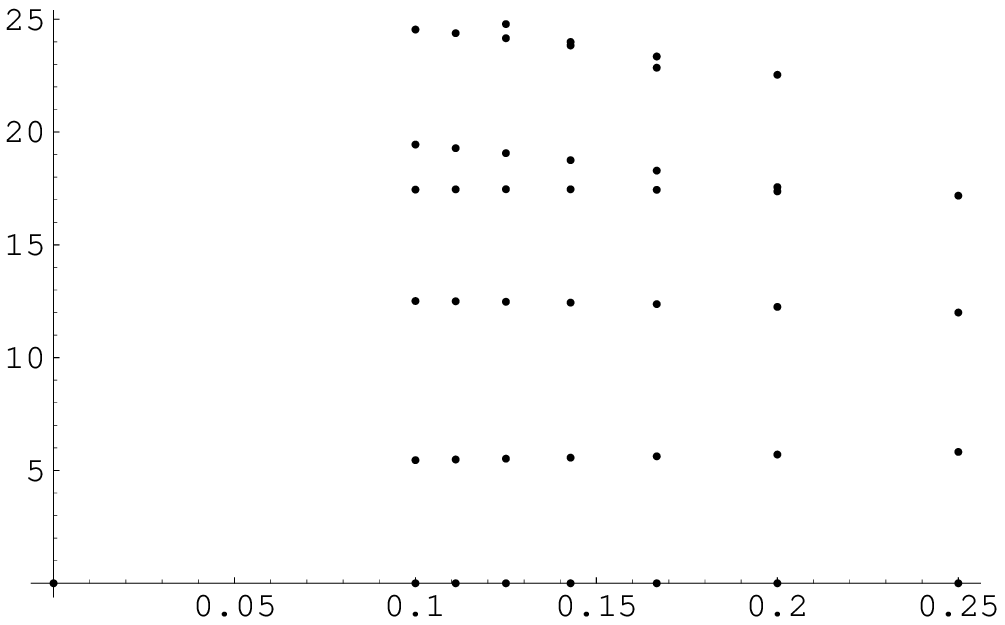}&
\epsfxsize=2.5in \epsffile{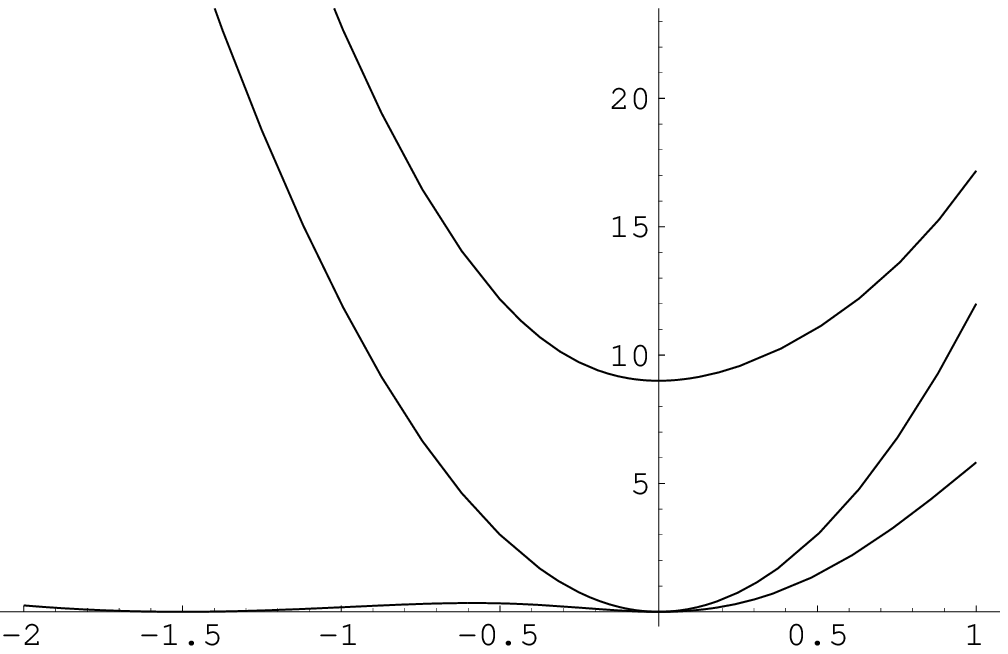}\\
(a)&(b)
\end{tabular}
\caption{Pure fermionic model.
(a) Mass spectrum in units of $g^2 N /\pi$ at $h=1$ as function of
$1/k$.
(b) Mass eigenvalues at $k=4$ as functions of coupling $h$.
\label{FermOnlySpec}}
\end{figure}

We begin with the pure fermionic system $\lambda\Psi$. This system has $(1,0)$
  supersymmetry and the supercharge has the form:
\be
Q^-=-\frac{g}{\sqrt{2}}\int dx^-\left(\left\{\lambda,\lambda\right\}+
2h\psi\psi^\dagger\right)
\frac{1}{\partial_-}\lambda.
\ee
After substituting the expansions (\ref{expandLambda}), (\ref{expandPsi})
one gets the mode decomposition of the supercharge:
\bea\label{FrmOnlSuchMode}
&&Q^-=\frac{i2^{-1/4}g\sqrt{L}}{\pi}\sum_{k_1=1}^\infty\sum_{k_2=1}^\infty
\left\{\frac{h}{k_1}\left[{\tilde d}^\dagger_i(k_2)
b^\dagger_{ij}(k_1){\tilde d}_j(k_1+k_2)+\right.\right.\\
&&\left.{\tilde d}^\dagger_j(k_1{+}k_2){\tilde d}_i(k_2)b_{ij}(k_1)+
b^\dagger_{ij}(k_1)
{d}^\dagger_j(k_2){d}_i(k_1{+}k_2)+
{d}^\dagger_i(k_1{+}k_2)b_{ij}(k_1){d}_j(k_2)
\right]+\nonumber\\
&&\left.\left(\frac{1}{k_1}+\frac{1}{k_2}-\frac{1}{k_1{+}k_2}\right)\left[
b^\dagger_{ik}(k_1)b^\dagger_{kj}(k_2)b_{ij}(k_1{+}k_2)+
b^\dagger_{ij}(k_1{+}k_2)b_{ik}(k_1)b_{kj}(k_2)\right]\right\}\nonumber
\eea
The mass spectrum of
this theory is presented in figure \ref{FermOnlySpec}a (we truncated it at
$M^2=25$ and chose $h=1$). There is one massless state and all other masses
converge to finite continuum limits. We
illustrate this convergence for the five lowest states in the
Table ~\ref{FermOnlyStateTable}. Using the structure of the supercharge
(\ref{FrmOnlSuchMode}), one can easily calculate the wavefunction of 
the massless
state for any finite resolution as well as its continuum limit. This state
appears to have only two partons in it and the wavefunction is equal to a
constant:
\be\label{FermOnlyMslsState}
|M=0,K\rangle=C\sum_{n=1}^{K-1}{\tilde 
d}^\dagger_i(n){d}^\dagger_i(K-n)|0\rangle
\rightarrow
\int_0^{P^+}dp\ {\widetilde D}^\dagger_i(p)D^\dagger_i(P^+-p)|0\rangle.
\ee
Here we have introduced the continuum modes $D^\dagger$ and
${\widetilde D}^\dagger$. One can see that the state with the wavefunction
(\ref{FermOnlyMslsState}) stays massless for an arbitrary values of coupling
$h$. It interesting to note that a massless state with constant wavefunction
was also observed in a non--supersymmetric adjoint QCD in two dimensions
\cite{anp97}.

\begin{figure}
\begin{tabular}{cc}
\epsfxsize=2.5in \epsffile{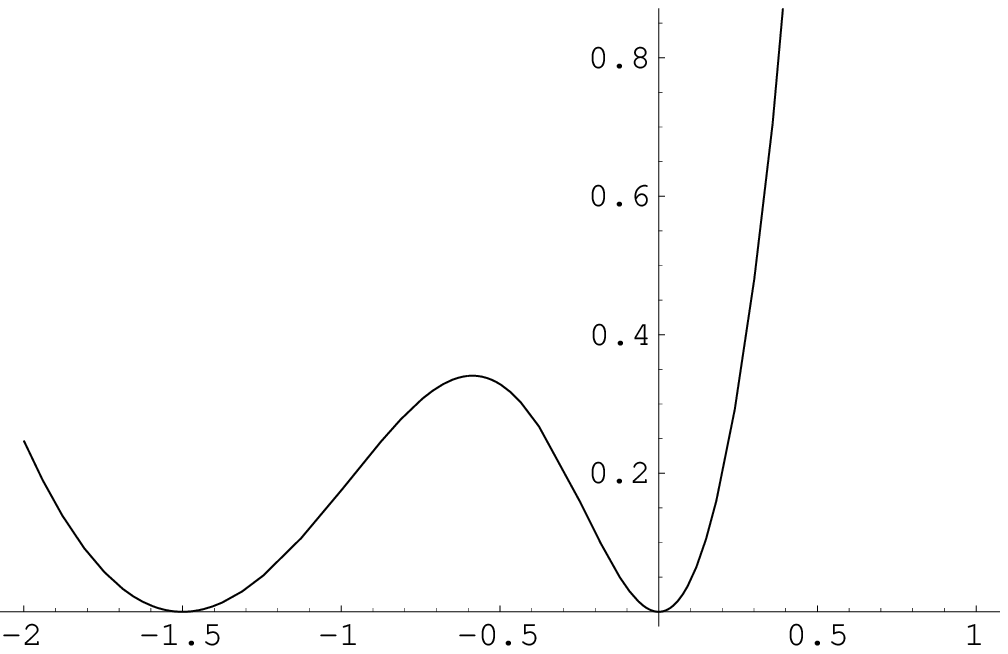}&
\epsfxsize=2.5in \epsffile{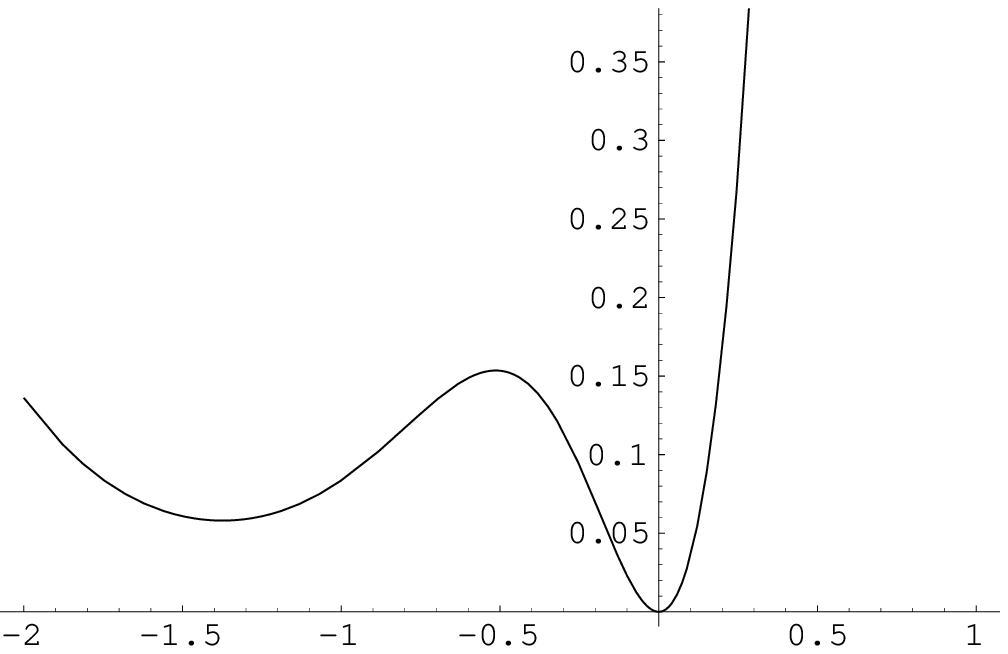}\\
(a)&(b)
\end{tabular}
\caption{Pure fermionic model: lowest nonzero mass as function of coupling
$h$ at resolutions $4$ (a) and $5$ (b).\label{StrangeState}
}
\end{figure}

\begin{figure}
\begin{tabular}{cccc}
\epsfxsize=1.2in \epsffile{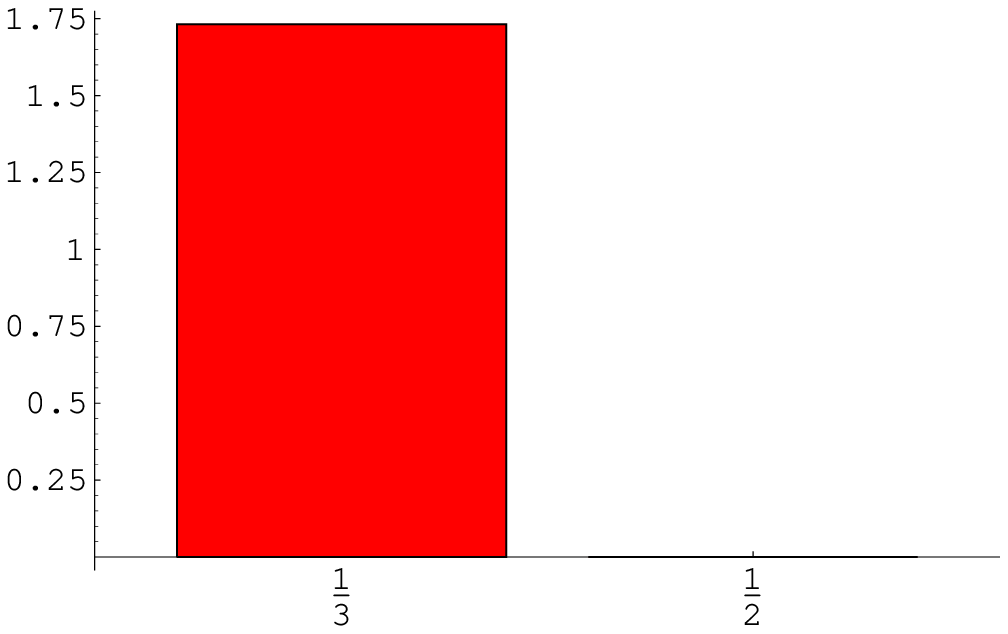}&
\epsfxsize=1.2in \epsffile{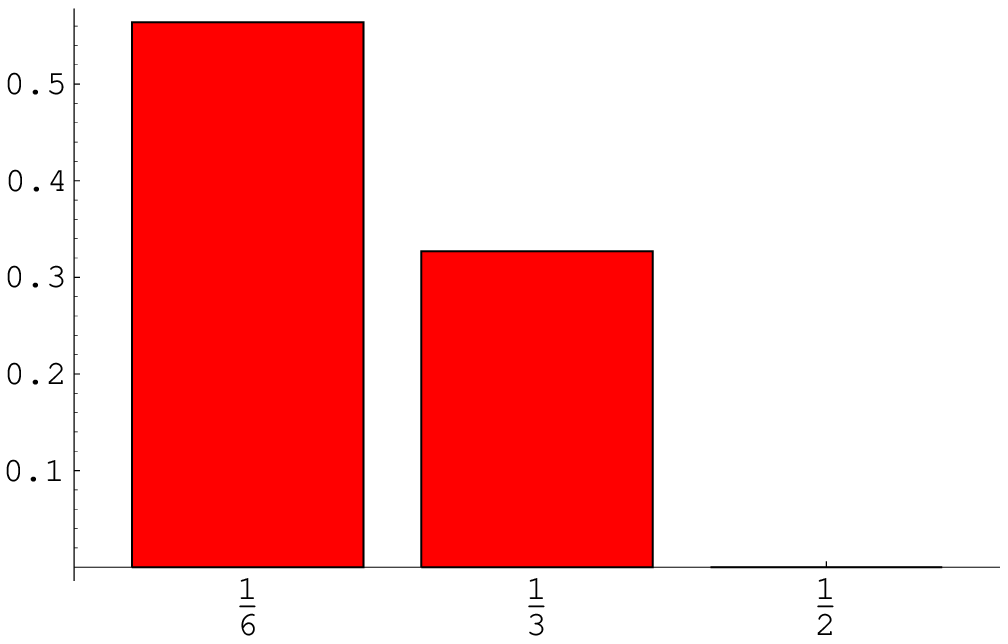}&
\epsfxsize=1.2in \epsffile{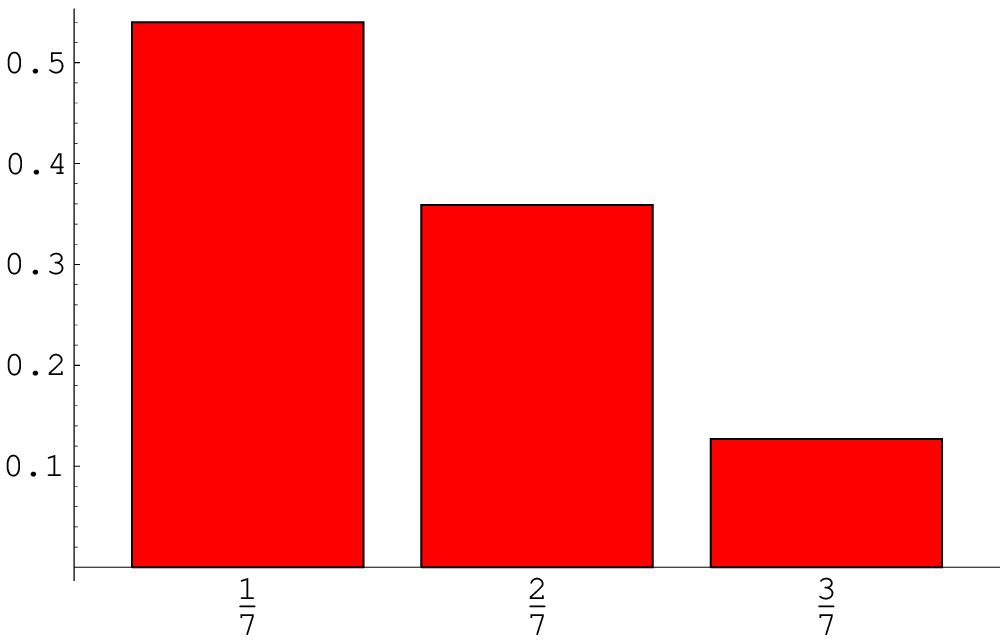}&
\epsfxsize=1.2in \epsffile{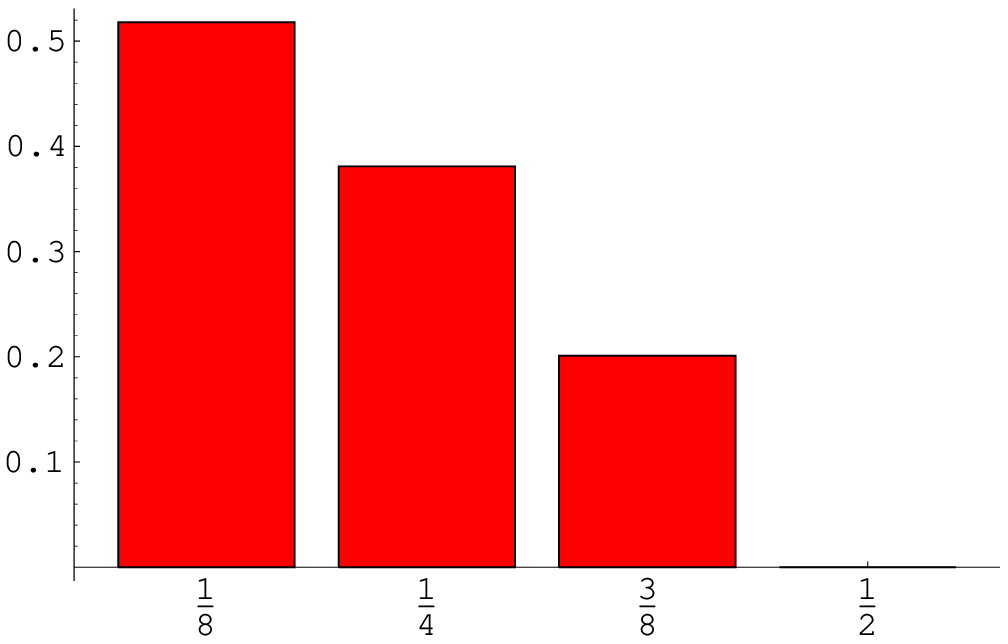}\\
(a)&(b)&(c)&(d)
\end{tabular}
\caption{Pure fermionic model: DLCQ wavefunction of antisymmetric massless
state appearing at the critical value of coupling. Only half of the
wavefunction ($p\le q$) is presented at resolutions $4$ (a), $6$ (b),
$7$ (c),$8$ (d).\label{FermOnlyWaveFunc}
}
\end{figure}

It is also interesting to look at other masses
as functions of the coupling constant. Figure \ref{FermOnlySpec}b shows this
dependence for the resolution $K=4$. We should note that this behavior is
typical for all values of resolution and for all of the systems we are
considering
in this section. In particular one can see that the lowest states stays near
$M^2=0$ for a wide range of negative couplings. A closer look at this
state at resolutions $4$ and $5$ is presented in figure \ref{StrangeState}.
Looking at higher resolutions, we observe, that this state become massless for
some value of coupling at all resolutions except $K=5$, thus the graph
\ref{StrangeState}a is typical, while \ref{StrangeState}b is an artifact of
the DLCQ. Interestingly, such odd behavior at $k=5$ is observed for all three
systems we are studying here. The values of the critical coupling and the
extrapolation to the continuum limit is presented in Table
\ref{FermOnlyCritCoup}. The wavefunction of this massless state is
concentrated in the two--parton sector: at resolutions $4$, $6$, $7$ and $8$
the two--particle sector contains $82\%$, $92\%$, $93\%$ and $95\%$ of
wavefunction,
however we believe that in continuum limit the wavefunction has small, but
nonzero contributions from sectors with an arbitrarily large number of partons.
This property is common for massless states in other supersymmetric theories
  \cite{alp98a}. While we were not able to solve for the continuum
wavefunction in two--parton sector, the SDLCQ results presented in figure
\ref{FermOnlyWaveFunc} points to a linear behavior in the continuum limit.
Note that for the state we are looking at:
\be
|M=0,K,h\rangle\approx \sum_{n=1}^{K-1}f(n,K-n){\tilde d}^\dagger_i(n)
{d}^\dagger_i(K-n)|0\rangle,
\ee
the wavefunction is antisymmetric: $f(p,q)=-f(q,p)$, so in figure
\ref{FermOnlyWaveFunc} we present only the region $p\le q$.

\begin{table}
\begin{center}
\caption{Pure fermionic model: lowest massive states.}
\label{FermOnlyStateTable}
\  \newline
\begin{tabular}{|c|c|c|c|c|c|c|c|c|}
\hline
state &K=4&  K=5&   K=6&    K=7&    K=8&    K=9&      K=10&    K=$\infty$\\
\hline
1&    5.82&  5.71&   5.63&  5.57&   5.52&   5.48&     5.46&    5.21  \\
\hline
2&    12&   12.25&   12.37& 12.44&  12.48&  12.50&    12.51&    12.90\\
\hline
3&   17.18& 17.37&   17.44&  17.46& 17.46&  17.46&     17.45&    17.67\\
\hline
4&    --- & 17.56&   18.29&  18.75& 19.06&  19.28&    19.45&   21.39\\
\hline
5&    --- & 22.54&   22.85&  23.84& 24.16&  24.38&   24.54&   26.80\\
\hline
\end{tabular}
\end{center}
\end{table}
\begin{table}
\begin{center}
\caption{Pure fermionic model: critical coupling as function of resolution.}
\label{FermOnlyCritCoup}
\  \newline
\begin{tabular}{|c|c|c|c|c|c|c|c|}
\hline
K   &K=4&  K=5&   K=6&    K=7&    K=8&    K=9&       K=$\infty$\\
\hline
h& -1.50&  ---& -1.48&  -1.59&  -1.35&  -1.27&  -1.25    \\
\hline
\end{tabular}
\end{center}
\end{table}

\begin{figure}
\begin{tabular}{cc}
\epsfxsize=2.5in \epsffile{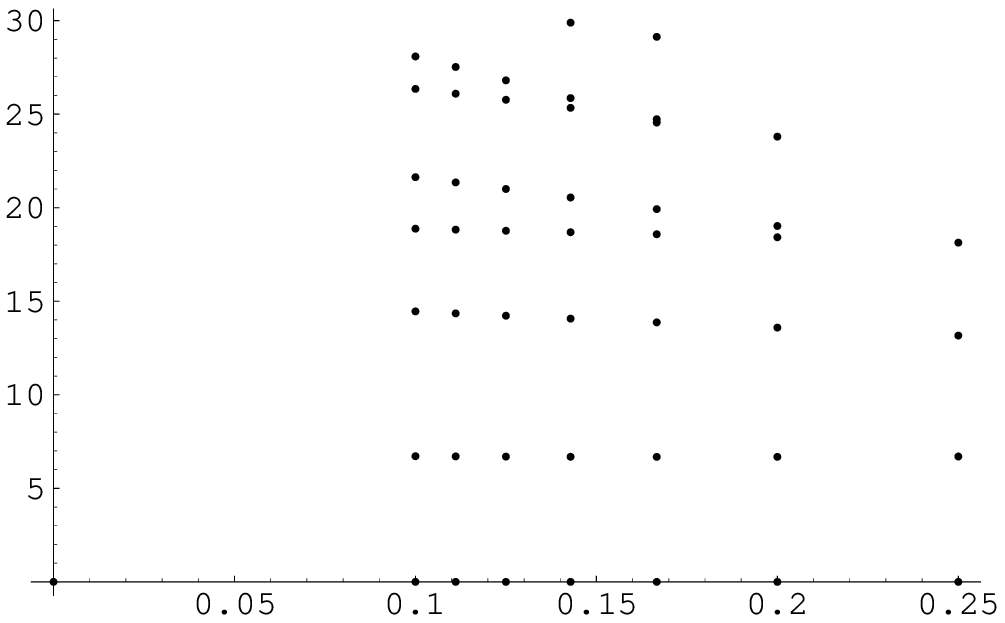}&
\epsfxsize=2.5in \epsffile{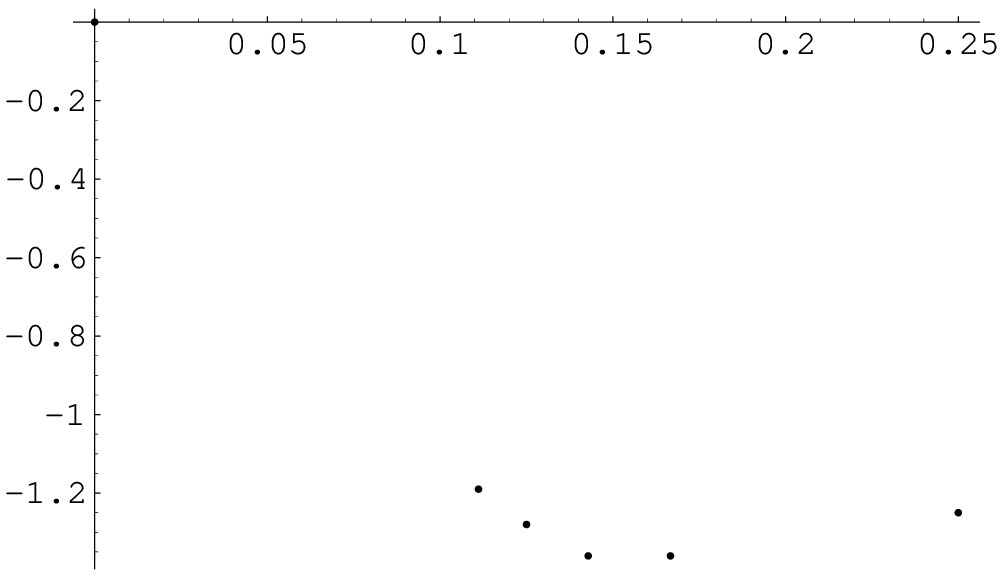}\\
(a)&(b)
\end{tabular}
\caption{$\lambda\xi$ model:
(a) Mass spectrum in units of $g^2 N /\pi$ at $h=1$ as function of
$1/k$.
(b) Critical coupling $h_{cr}$ as function of $1/k$.
\label{XiLaSpec}}
\end{figure}

Let us now discuss two other theories without adjoint scalars. As we already
mentioned, their properties are similar to the ones of the $\lambda\psi$
system,
so we will consider both $\lambda\xi$ and $\lambda\psi\xi$ models only
briefly. The $\lambda\xi$ system has $(1,0)$ supersymmetry and its
supercharge reads:
\be
Q^-=-\frac{g}{\sqrt{2}}\int dx^-\left(\left\{\lambda,\lambda\right\}+
i\sqrt{2}h\xi\partial_-\xi^\dagger-i\sqrt{2}h\partial_-\xi\xi^\dagger\right)
\frac{1}{\partial_-}\lambda.
\ee
The mass spectrum of this model is presented in figure \ref{XiLaSpec}a
(again, we put $h=1$) and the extrapolation of the masses to the continuum
limit is given in Table \ref{XiLaTable}. The masses exhibit the same
coupling
dependence as in the pure fermionic model (the only exception is the absence
of the coupling--independent massless state), in particular there exists a
critical coupling at which one of the states becomes massless. The values of
this critical coupling are plotted in figure \ref{XiLaSpec}b and
for the highest resolution we have $h_{cr}=-1.2$.

Finally we analyze the system which includes everything except for the
adjoint scalar ($\lambda\psi\xi$ in our notation). This model has two
supercharges, but we look only at one of
them:
\be
Q^-=-\frac{g}{\sqrt{2}}\int dx^-\left(\left\{\lambda,\lambda\right\}+
i\sqrt{2}h\xi\partial_-\xi^\dagger-i\sqrt{2}h\partial_-\xi\xi^\dagger+
2h\psi\psi^\dagger\right)
\frac{1}{\partial_-}\lambda.
\ee
The second supercharge can be formally constructed:
\be
Q^+=2\int dx^-\left(
\frac{i}{2}\partial_-\xi^\dagger\psi-\frac{i}{2}\psi^\dagger\partial_-\xi-
\frac{i}{2}\xi^\dagger\partial_-\psi+\frac{i}{2}\partial_-\psi^\dagger\xi
\right),
\ee
but its square does not give the canonical $P^+$ but rather:
\be
\left\{Q^+,Q^+\right\}=2\sqrt{2}P^+-2\sqrt{2}\int dx^-\
\lambda\partial_-\lambda.
\ee
Moreover, from this expression one concludes that
\be
\left[\left\{Q^+,Q^+\right\},Q^-\right]\ne 0,
\ee
thus the two supercharges do not anticommute and they cannot be diagonalized
simultaneously. Our formulation of the bound state problem is based on
diagonalization of $P^+$ and $\left(Q^-\right)^2$, which still commute,
and the $Q^+$ operator must be abandoned.

\begin{figure}
\begin{tabular}{cc}
\epsfxsize=2.5in \epsffile{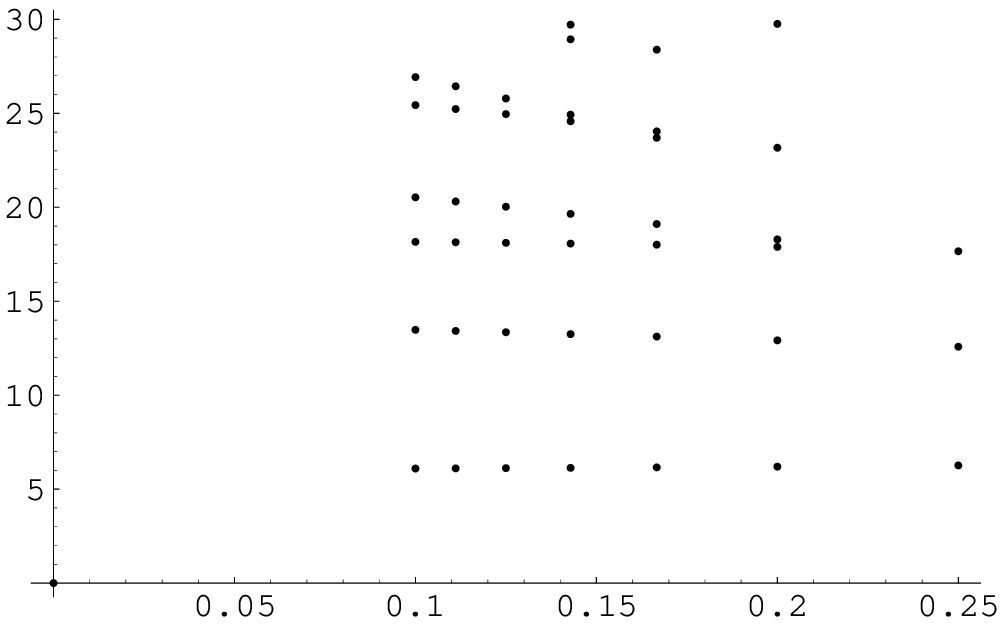}&
\epsfxsize=2.5in \epsffile{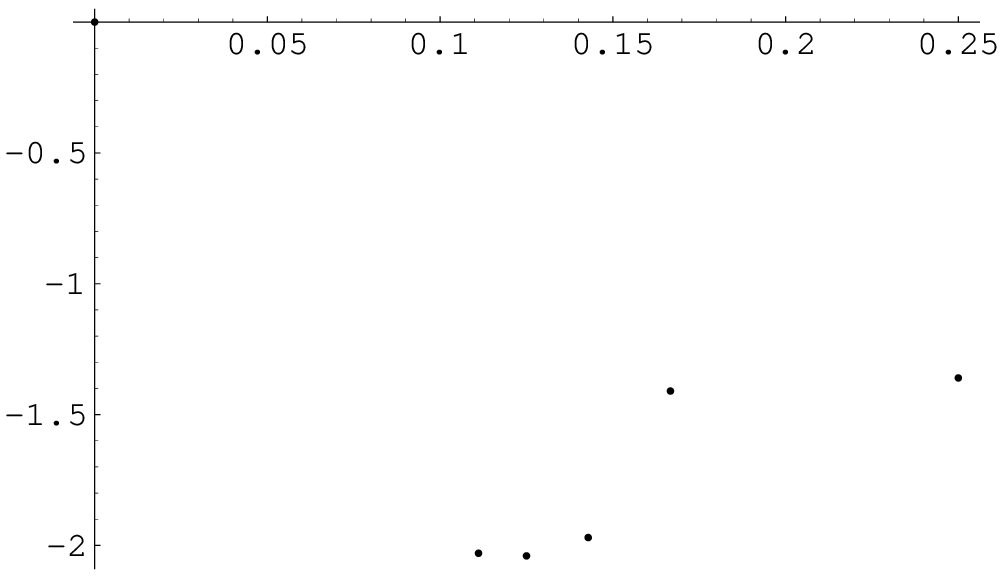}\\
(a)&(b)
\end{tabular}
\caption{$\lambda\psi\xi$ model:
(a) Mass spectrum in units of $g^2 N /\pi$ at $h=1$ as function of
$1/k$.
(b) Critical coupling $h_{cr}$ as function of $1/k$.
\label{NoGaugeSpec}}
\end{figure}

In the large $N$ calculations there are four different sectors to consider.
One can start from either one of the four types of states:
\be
{\tilde d}^\dagger b^\dagger\dots b^\dagger {d}^\dagger|0\rangle,\quad
{\tilde c}^\dagger b^\dagger\dots b^\dagger c^\dagger|0\rangle,\quad
{\tilde c}^\dagger b^\dagger\dots b^\dagger {d}^\dagger|0\rangle,\quad
{\tilde d}^\dagger b^\dagger\dots b^\dagger {c}^\dagger|0\rangle,\quad
\ee
then $Q^-$ acts only inside the corresponding subspace.
The first and second sectors
reproduce the results we just obtained for $\lambda\psi$ and
$\lambda\xi$ models. Two remaining models are mapped
into each other under the Z2 transformation:
\be
b_{ij}\leftrightarrow b_{ji},\quad {\tilde c}_i\leftrightarrow c_i,\quad
{\tilde d}_i\leftrightarrow d_i,
\ee
which is a symmetry of $Q^-$. The spectrum for one of these sectors is
presented in figure \ref{NoGaugeSpec}a and in the table \ref{NoGaugeTable}
and
the critical coupling is plotted in figure \ref{NoGaugeSpec}b. Note that we
don't have a coupling--independent massless state in this sector.
The critical coupling for resolutions $K>6$ is relatively constant at
$h_{cr}=-2.0$.
%
\begin{table}
\begin{center}
\caption{$\lambda\xi$ model: lowest massive states.}\label{XiLaTable}
\  \newline
\begin{tabular}{|c|c|c|c|c|c|c|c|c|}
\hline
state &K=4&  K=5&   K=6&    K=7&    K=8&    K=9&      K=10&    K=$\infty$\\
\hline
1&    6.70& 6.68&  6.68&   6.69&   6.70&   6.70&      6.72&    6.71 \\
\hline
2&  13.16& 13.60& 13.87&  14.07&   14.23& 14.35&     14.46&    15.30 \\
\hline
3& 18.13& 18.42&  18.58&  18.69&   18.77& 18.83&     18.88&    19.20 \\
\hline
4&  --- & 19.02&  19.93&  20.55&   21.00& 21.35&     21.63&    23.96\\
\hline
5&  --- & 23.80&  24.55&  25.34&   25.77& 26.09&     26.35&    29.00\\
\hline
\end{tabular}
\end{center}
\end{table}
\begin{table}
\begin{center}
\caption{$\lambda\psi\xi$ model: lowest massive states.}\label{NoGaugeTable}
\  \newline
\begin{tabular}{|c|c|c|c|c|c|c|c|c|}
\hline
state &K=4&    K=5&    K=6&    K=7&    K=8&    K=9&      K=10&    K=$\infty$\\
\hline
1&    6.26&   6.20&   6.16&   6.13&   6.11&   6.10&      6.11&  5.98\\
\hline
2&   12.58&  12.92&  13.12&  13.25&  13.35&  13.42&     13.48&  14.09\\
\hline
3&   17.66&  17.89&  18.01&  18.07&  18.11&  18.14&     18.16&  18.50\\
\hline
4&    --- &  18.29&  19.11&  19.65&  20.03&  20.31&     20.53&  22.80\\
\hline
5&  --- & --- &      23.70&  24.58&  24.96&  25.23&     25.44&  31.30\\
\hline
\end{tabular}
\end{center}
\end{table}

\section{Models containing adjoint bosons.}
\label{SecGauge}
\begin{figure}
\begin{tabular}{cc}
\epsfxsize=2.5in \epsffile{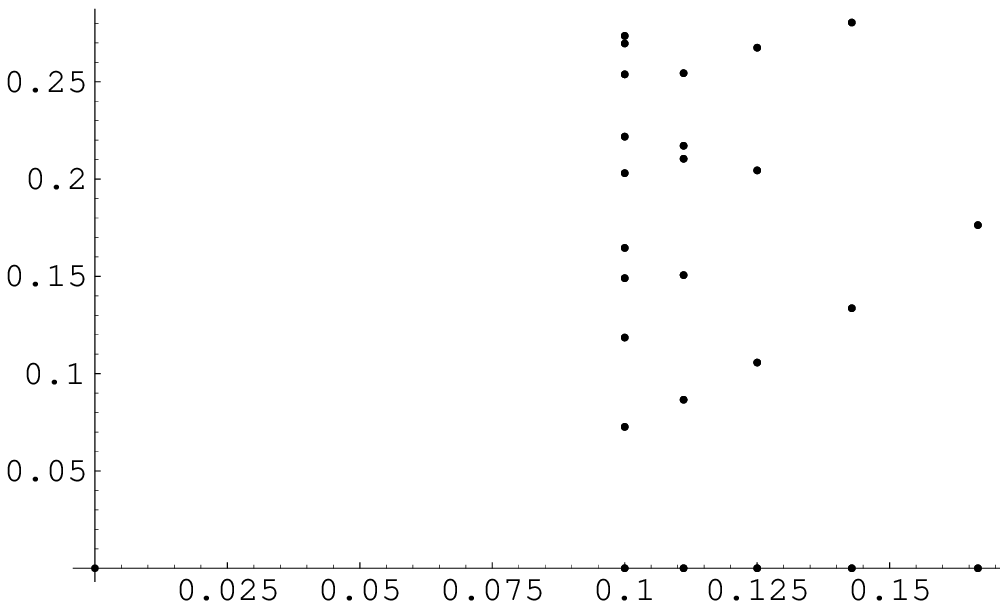}&
\epsfxsize=2.5in \epsffile{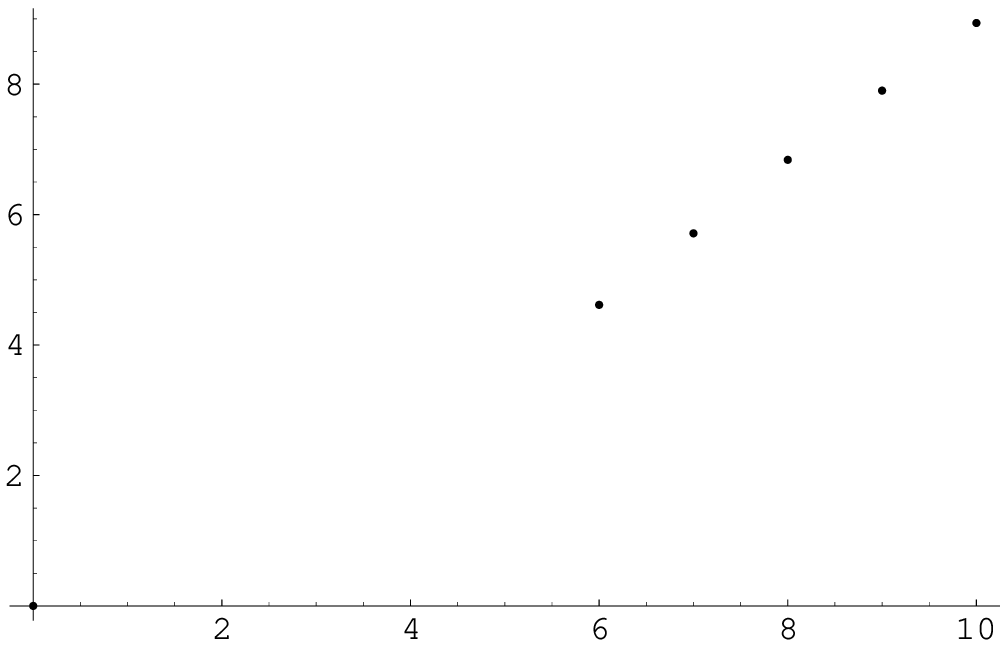}\\
(a)&(b)
\end{tabular}
\caption{$A\psi\xi$ model:
(a) Mass spectrum in units of $(g')^2 N /\pi$ as function of
$1/k$.
(b) Maximum mass in units of $(g')^2 N /\pi$ as function of resolution. }
\label{FigAPX}
\end{figure}

\begin{figure}
\begin{tabular}{cc}
\epsfxsize=2.5in \epsffile{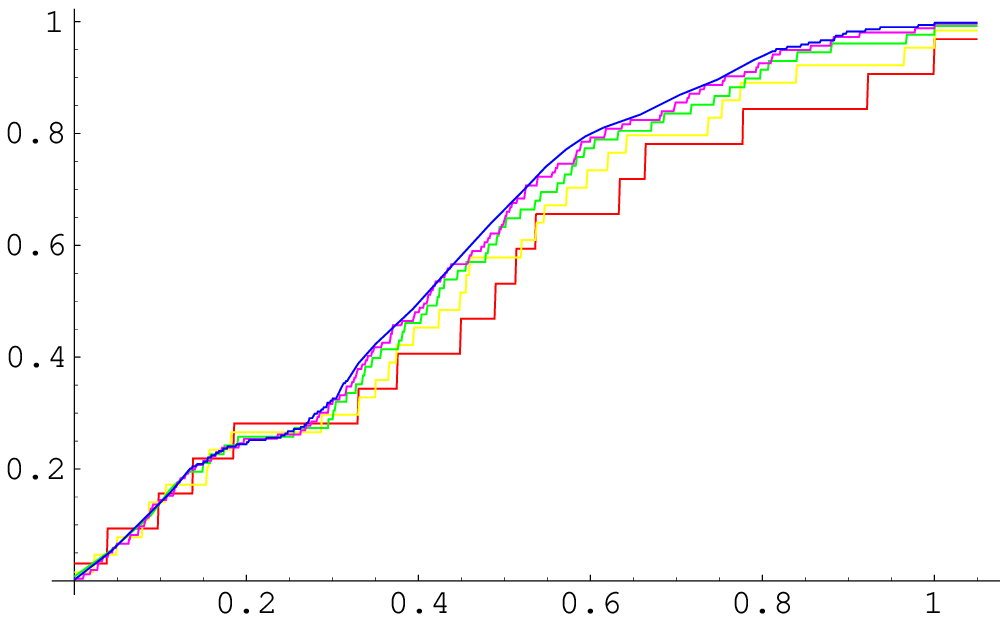}&
\epsfxsize=2.5in \epsffile{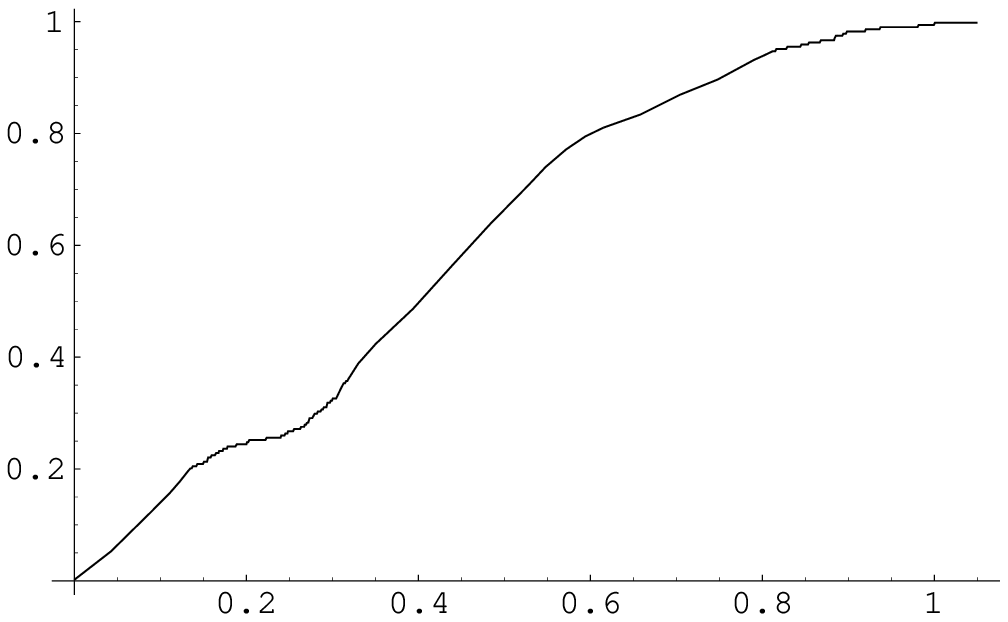}\\
(a)&(b)
\end{tabular}
\caption{$A\psi\xi$ model:
(a) Color plot of function $F(x)$ at resolutions $k=6\dots 10$,
(b) The same function at $K=10$ as a best available approximation to the
continuum limit.}
\label{FigAPXall}
\end{figure}

Let us now discuss the models with an adjoint scalar $A^2$. We will 
see that the
general property of such theories is the presence of a large number of low
energy bound states which complicates the
extrapolation to the continuum  limit. We begin with the system which does not
have a gluino as a dynamical
field. Although this model has two supercharges:
\bea
Q^-=-2g'\int dx^-\left(\xi^\dagger A^2\psi+\psi^\dagger A^2\psi\right),\\
Q^+=2\int dx^-\left(
\frac{i}{2}\partial_-\xi^\dagger\psi-\frac{i}{2}\psi^\dagger\partial_-\xi-
\frac{i}{2}\xi^\dagger\partial_-\psi+\frac{i}{2}\partial_-\psi^\dagger\xi
\right),
\eea
they do not commute, thus we will ignore the supercharge $Q^+$ in our
consideration (this situation is analogous to the case of
$\lambda\xi\psi$ model). We constructed mesonic spectrum of this model and the
  lowest masses are presented in figure \ref{FigAPX}a. One can see that the
number of light states grows with resolution, in fact we will argue that in
the  continuum limit this model has a continuous mass spectrum. But first let
us
look at the heaviest bound state one can construct at a given value of
resolution. The masses of such states are plotted in figure \ref{FigAPX}b, and
one can see that this graph has a good linear approximation:
\be
M^2_{max}(K)=\frac{(g')^2 N}{\pi}\left(1.08K-1.86\right),
\ee
where $K$ is a value of resolution. The negative constant in the above
expression is not important, since one cannot consider $K<4$ and the
interesting limit is $K\rightarrow\infty$. For the total number of bosonic
states at a particular resolution, simple combinatorics gives:
\be
N_{total}(K)=2^{K}.
\ee
In order to analyze the continuum
limit of the spectrum we will study the density of states $dN/N_{total}(K)$ as
a function of the reduced mass: $dM^2/M^2_{max}(K)$. In particular we
will plot the distribution function defined as
\be
F(M^2/M^2_{max}(K))=\frac{N({\mbox{Mass}}<M)}{N_{total}}.
\ee
Such plots for different resolutions are presented in color in figure
\ref{FigAPXall}a, which gives a convincing argument for the convergence of the
function $F(x)$ in the continuum limit. In figure \ref{FigAPXall}b we
present this function for resolution $10$, which is a good approximation to
the continuum limit.


\begin{figure}
\begin{tabular}{cc}
\epsfxsize=2.5in \epsffile{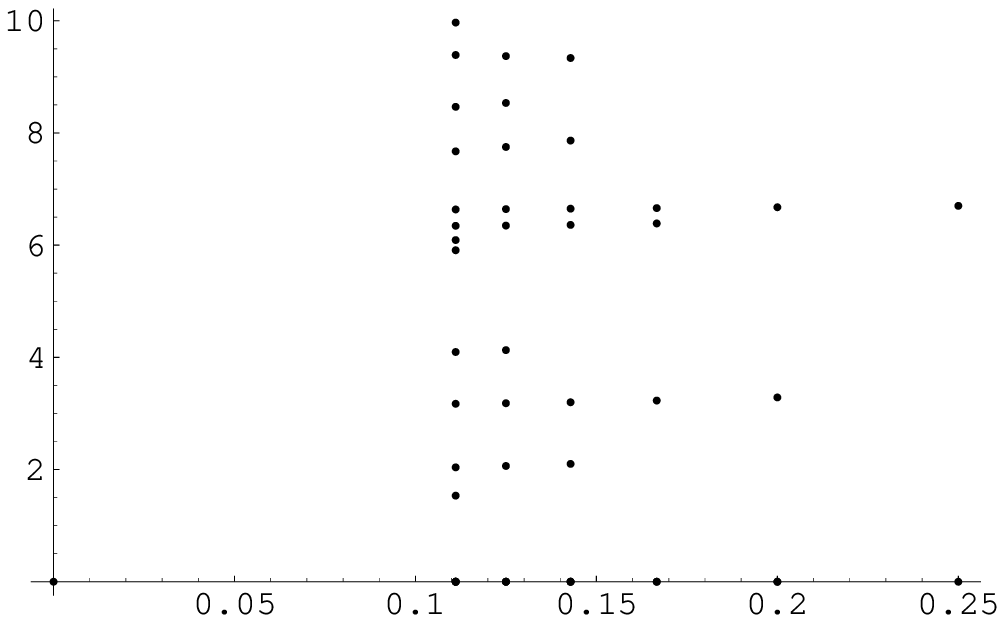}&
\epsfxsize=2.5in \epsffile{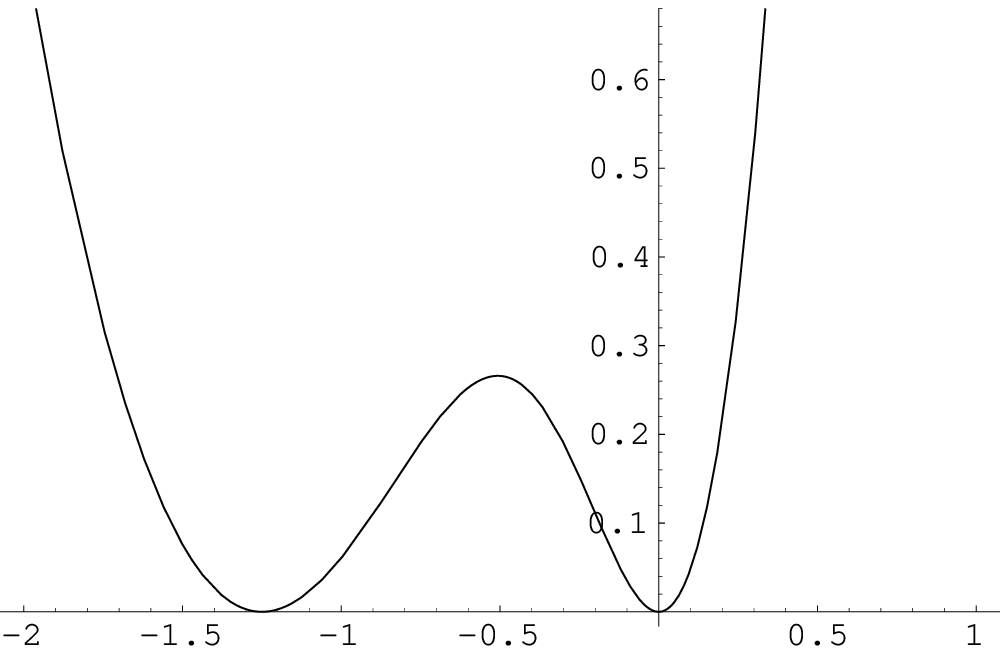}\\
(a)&(b)
\end{tabular}
\caption{$A\lambda\psi$ model:
(a) Mass spectrum at $h=1$ in units of $(g')^2 N /\pi$ as function of
$1/k$.
(b) Lowest nonzero mass at $K=4$ as function of coupling $h$.}
\label{FigALP}
\end{figure}

Let us now discuss the models that include both fields in the adjoint
representation. First we consider the $A\lambda\psi$ system. It has two
supercharges which don't commute, so we look only at one of them:
\be
Q^-=-2g\int dx^-\left(i[A^2,\partial_-A^2]+
\frac{1}{\sqrt{2}}\{\lambda,\lambda\}+\sqrt{2}h\psi\psi^\dagger
\right)\frac{1}{\partial_-}\lambda.
\ee
The mass spectrum, obtained as the result of the diagonalization of
Hamiltonian $P^-=\left(Q^-\right)^2$, is presented in the figure
\ref{FigALP}a. As usual we put $h=1$ and truncated the spectrum at some value
of mass (in this case $M^2=10$). The states can
be easily traced from one resolution to another, and a new low state appears at
every even resolution and a new massless state appears at every odd
resolution
(thus there is one massless state at $K=4$, two massless states at $K=5$
and $K=6$, $3$ massless states at $K=7,8$ and so on). Such behavior was
observed in other supersymmetric systems with adjoint scalars \cite{alp2} and
it points to a continuum spectrum in the limit $K=\infty$. One can also look
at the coupling dependence of the states, in particular the lowest state with
nonzero mass at $K=4$ becomes massless at the coupling $h=-1.25$ (see figure
\ref{FigALP}b). But since
this state ultimately becomes a part of a continuous spectrum, its properties
are not as interesting as its counterpart in the $\lambda\psi$ model.

\begin{figure}
\begin{tabular}{cc}
\epsfxsize=2.5in \epsffile{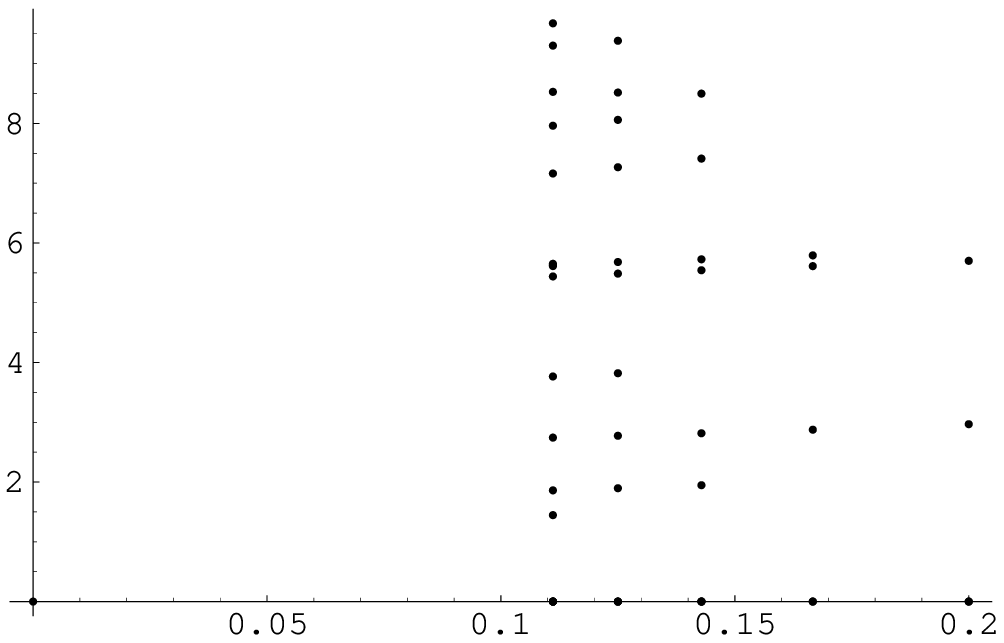}&
\epsfxsize=2.5in \epsffile{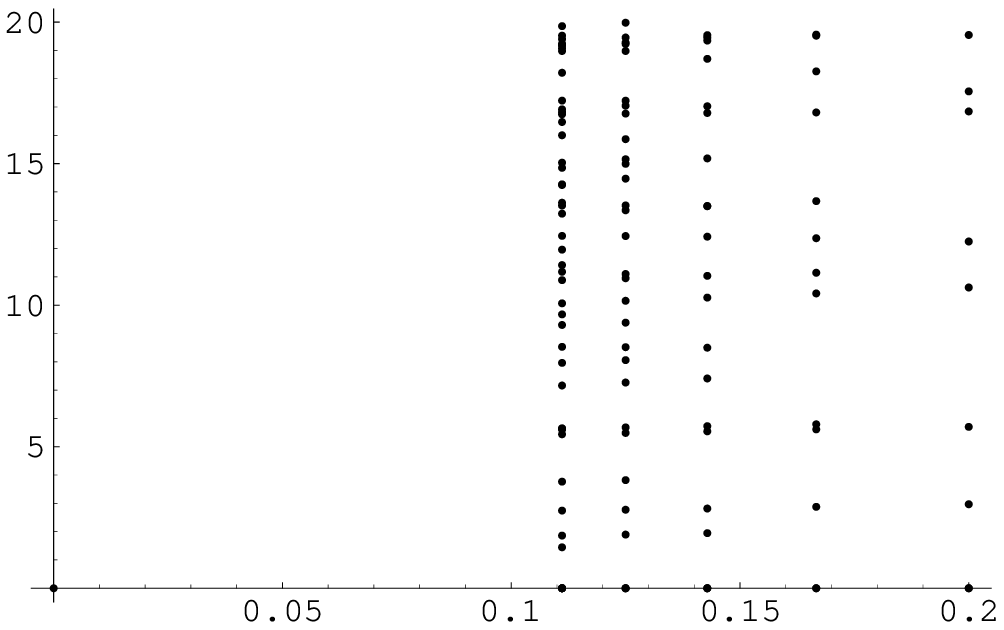}\\
(a)&(b)
\end{tabular}
\caption{$A\lambda\xi$ model: Mass spectrum at $h=1$ in units of
$(g')^2 N /\pi$ as function of $1/k$.}
\label{FigALX}
\end{figure}

As we saw in our study of the models without gauge fields, the theories which
differ only by replacing $\psi$ by $\xi$, behave in the same fashion. The
same is true for the models with adjoint scalar. The spectrum of the 
$A\lambda\xi$
model is presented in figure \ref{FigALX}, it also converges to a continuous
spectrum and has a critical coupling $h$ at any $K\ne 5$. The figure
\ref{FigALX}b illustrates that it is not only low mass states that appear at
high resolution, but all values of mass seem to be filled in the continuum
limit.

\begin{figure}
\epsfxsize=3.5in \epsffile{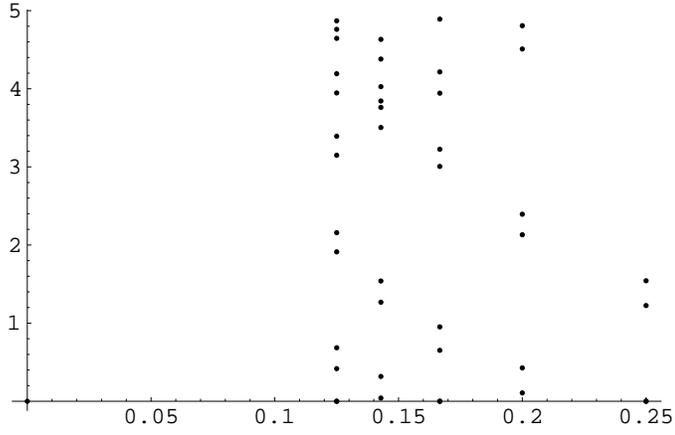}
\caption{Complete two dimensional model:
Mass spectrum at $h=1$ in units of $g^2 N /\pi$ as function of
$1/k$. }
\label{FigAllSpec}
\end{figure}

\begin{figure}
\begin{tabular}{cc}
\epsfxsize=2.5in \epsffile{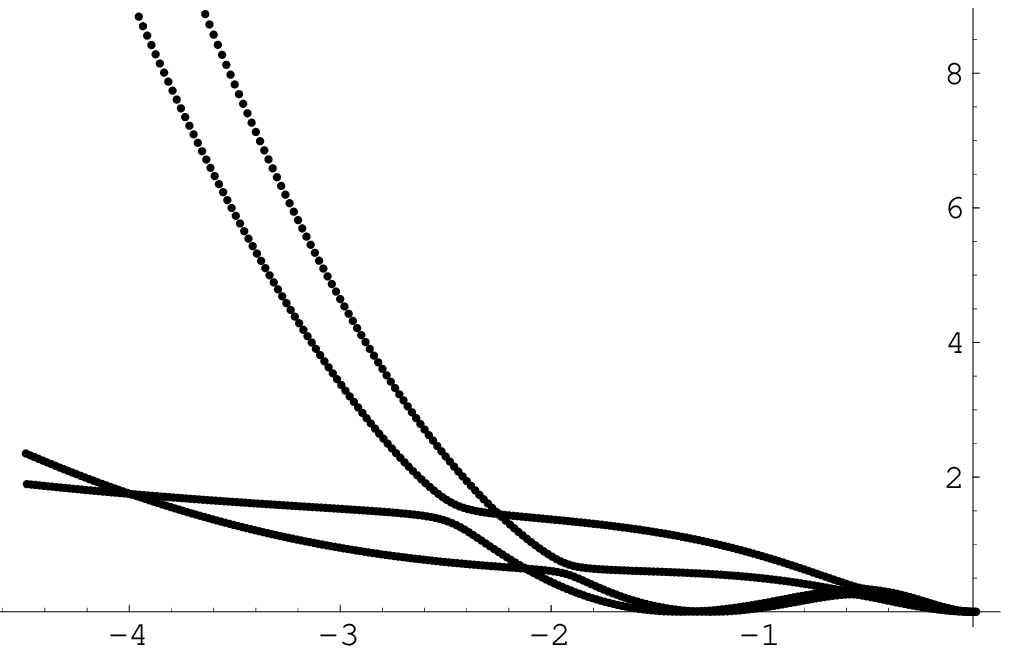}&
\epsfxsize=2.5in \epsffile{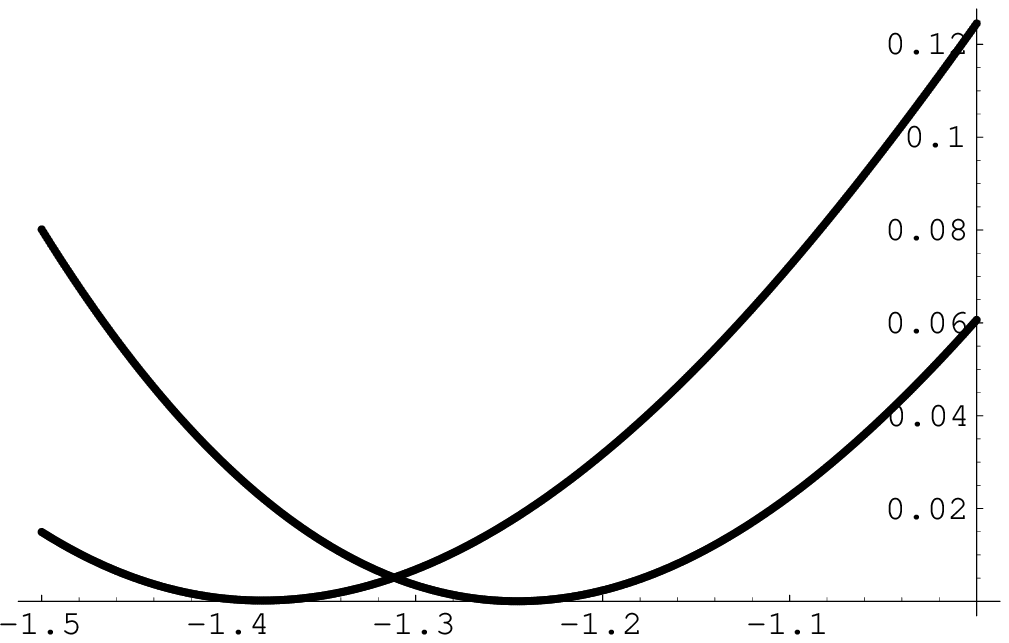}\\
(a)&(b)
\end{tabular}
\caption{Complete two dimensional model: (a) Lowest masses at
resolution $4$ as functions of coupling $h$, (b) Lowest
mass at resolution $4$ near critical couplings.}
\label{FigAllCoupl}
\end{figure}

Finally we consider the system without truncation, i.e. we study the
$A\lambda\psi\xi$ model. Unlike all other systems we studied in this section,
it has a complete $(1,1)$ supersymmetry. Thus the two supercharges given by
(\ref{Qplus}) and (\ref{Qminus}) anticommute and can be diagonalized
simultaneously. Thus, apart from the massless sector, the spectrum is 
four--fold
degenerate and we can look only at a quarter of the theory, while diagonalizing
the mass operator. In particular, we consider only bosonic states with an
even number of creation operators $a^\dagger$. The combined action of $Q^+$
and $Q^-$ gives a boson with an odd number of creation operators $a^\dagger$,
while the action of either one of the supercharges leads to the fermionic
sectors. The low energy spectrum in a single sector at $h=1$ is presented in
figure \ref{FigAllSpec}, there are two massless states for every even value of
resolution. One can see that the presence of these states is not the only
difference between
the odd and even values of $K$, it seems that we are dealing with the 
SDLCQ of two
different theories. Of course, in the large $K$ limit they should converge to
the same result, and, as figure \ref{FigAllSpec} demonstrates, the resulting
theory has many light states and the possibility of a continuum spectrum is
not ruled out. Note that the lowest states appear in almost degenerate
pairs. The explanation of this doubling poses an interesting question, which
might be answered by performing a careful study of the wavefunctions. We also
looked at the lowest masses as functions of a coupling $h$, and the result for
resolution $4$ is presented in figure \ref{FigAllCoupl}. This shows two
peculiar properties of this model. First, there is a smooth interchange
between almost degenerate states, as opposed to the level crossing we observed
in the other theories. In addition to this, the model exhibits two critical
couplings at resolution $4$ and it is interesting to see whether this property
persists at higher resolutions. Unfortunately this model is the hardest one
to study since it has a maximal number
of fields. Also additional calculational effort is require to handle 
the fact that
the $A\lambda\psi\xi$ model
({\it{and only this  model}}) has a complex supercharge. We
hope to overcome both difficulties in the future work. In spite of these
difficulties we can already conclude that the $A\lambda\psi\xi$ has many
light states in the continuum limit and possibly it converges to a continuous
spectrum as the other models with adjoint scalars.
\section{Discussion.}

In this work we studied the mesonic mass spectrum of various supersymmetric
models. The calculations were performed in the framework of supersymmetric
DLCQ, namely we compactified the light--like coordinate $x^-$ on a finite
circle and performed a numerical diagonalization of supercharge $Q^-$.
We found that the systems with adjoint scalars tend to give a
continuous spectrum in the decompactification limit. For one of these systems
($A\psi\xi$ model) we found a limiting form of the distribution function for
the mass. By contrast, the models without adjoint bosons have well
defined spectra of bound states in the continuum limit. The only exception
from this rule is the system containing only the gaugino field, which seems not
to have any finite mass mesons in the decompactification limit. For 
the well defined
systems we found the masses of the lightest
mesons and demonstrated the fast convergence of SDLCQ approximation.

We also looked at the mass spectrum at different values of coupling constant.
The nontrivial phase  diagram is an essential property which distinguish the
models we considered here from the two dimensional systems studied previously
\cite{kutasov,sakai}. The coupling constant of a pure gauge theory in two
dimensions has a dimension of mass, thus all the bound state masses scale
like $g$, leaving no space for nontrivial coupling constant dependence. One
can avoid this problem by introducing the masses for a gauge field or its
superpartner, but such terms usually lead to breaking of either a gauge
invariance or supersymmetry\footnote{Actually the pure fermionic system
$\lambda$ is supersymmetric only at nonzero value of a particle mass
$m=g^2N/\pi$, but this still does not leave a space for an adjustable mass
parameter.}. Another way of introducing a free parameter in the theory is to
add a new supermultiplet with a different charge $g'$. Of course, this
parameter is not completely free: the quantization of charge requires the
value $g'/g$ to  be a rational number, but formally we can study the bound 
states
as functions of $h=g'/g$. We found an interesting property of the lowest
nonzero mass: it vanishes at a particular negative value of $h$. The nature
of this property is still unclear. One can also introduce the true free
parameter by considering the massive matter supermultiplet (now there is no
obstacle coming from gauge invariance) and we leave this possibility for a
future investigation.

\end{document}